\documentclass[10pt,twocolumn,letterpaper]{article}
\usepackage[accsupp]{axessibility}
\usepackage[pagenumbers]{wacv}      

\usepackage{graphicx}
\usepackage{amsmath}
\usepackage{amssymb}
\usepackage{amsthm}
\theoremstyle{definition}

\usepackage{booktabs}

\usepackage[pagebackref,breaklinks,colorlinks]{hyperref}

\usepackage[capitalize]{cleveref}
\crefname{section}{Sec.}{Secs.}
\Crefname{section}{Section}{Sections}
\Crefname{table}{Table}{Tables}
\crefname{table}{Tab.}{Tabs.}
\usepackage{multirow}

\def\wacvPaperID{434} 
\def\confName{WACV}
\def\confYear{2024}

\usepackage[utf8]{inputenc}
\usepackage{bbm}
\usepackage[table,xcdraw]{xcolor}
\usepackage{tabularx}
\renewcommand\L{\mathcal{L}}
\newcommand\blank{\mbox{-}}
\renewcommand\check{\checkmark}

\begin{document}
\title{Dynamic Multimodal Information Bottleneck for Multimodality Classification}

\author{Yingying Fang$^{1*}$, Shuang Wu$^3$\thanks{Equal Contribution}, Sheng Zhang$^1$, Chaoyan Huang$^4$, Tieyong Zeng$^4$, Xiaodan Xing$^2$, \\Simon Walsh$^1$, Guang Yang$^{1,2}$\thanks{Corresponding Author}\\
\\ $^1$\small{National Heart and Lung Institute, Imperial College London, London, SW7 2AZ, UK} \\
$^2$\small{Bioengineering Department and Imperial-X, Imperial College London, London, W12 7SL, UK}\\
$^3$\small{Black Sesame Technologies, Fusionopolis, Singapore} \\
$^4$\small{The Chinese University of Hong Kong, Shatin, Hong Kong}\\
{\tt\small y.fang;sheng.zhang;x.xing;s.walsh;g.yang@imperial.ac.uk}\\
{\tt\small wushuang@outlook.sg}\quad
{\tt\small cyhuang;zeng@math.cuhk.edu.hk}
}


\maketitle
\begin{abstract}
Effectively leveraging multimodal data such as various images, laboratory tests and clinical information is becoming increasingly attractive in a variety of AI-based medical diagnosis and prognosis tasks. Most existing multi-modal techniques only focus on enhancing their performance by leveraging the differences or shared features from various modalities and fusing feature across different modalities. These approaches are generally not optimal for clinical settings, which pose the additional challenges of limited training data, as well as being rife with redundant data or noisy modality channels, leading to subpar performance. 
To address this gap, we study the robustness of existing methods to data redundancy and noise and propose a generalized dynamic multimodal information bottleneck framework for attaining a robust fused feature representation. Specifically, our information bottleneck module serves to filter out the task-irrelevant information and noises in the fused feature, and we further introduce a sufficiency loss to prevent dropping of task-relevant information, thus explicitly preserving the sufficiency of prediction information in the distilled feature. We validate our model on an in-house and a public COVID19 dataset for mortality prediction as well as two public biomedical datasets for diagnostic tasks. Extensive experiments show that our method surpasses the state-of-the-art and is significantly more robust, being the only method to remain performance when large-scale noisy channels exist. Our code is publicly available at \url{https://github.com/ayanglab/DMIB}.
\end{abstract}

\section{Introduction}
Medical practitioners utilize various sources of data such as electronic health records, laboratory tests, genetic information and medical imaging modalities such as Computerized Tomography (CT), Magnetic Resonance Imaging (MRI), Positron Emission Tomography (PET) etc. for medical diagnosis and prognosis. Historically, partly due to domain gaps and specialization, clinicians primarily operated on each data modality in silo, drawing upon the distinctive features from an individual source to perform diagnosis and prognosis. Nonetheless, it is undeniable that integrating multi-modal data effectively would be beneficial for diagnosis and prognosis tasks, by not only incorporating extra guidance but also possibly providing novel insights, enabled by a more holistic understanding of the entirety. 

The advent of deep learning has sparked several lines of works centered on healthcare applications \cite{yu2018artificial}, among which AI-based medical image analysis \cite{shen2017deep} has enjoyed prominent success, with deep learning models achieving performance on par with or even surpassing radiologists on some tasks. Likewise, deep learning has also demonstrated immense potential for analysing electronic health records \cite{rajkomar2018scalable} and genetic information \cite{zeng2021deep}. Yet, despite impressive performance on individual modalities, developing techniques to effectively leverage multiple modalities remains challenging \cite{liang2022foundations,cai2019survey,muhammad2021comprehensive,behrad2022overview}.

The general multimodal learning approach is to separately train a model for each modality to extract a modal-specific feature vector, and subsequently fusing these individual features to obtain a multimodal feature representation. The fused feature is then propagated to downstream task modules to perform supervised learning. Common approaches for feature fusion include concatenation, attention-weighted, common subspace projection, graph-based, and transformer-based methods \cite{cui2022deep,acosta2022multimodal}.

While experimenting with these different fusion schemes in the context of clinical applications, we observe that they generally suffer from significant drops in model performance. Clinical applications typically have limited training data, and the difficulty is compounded by the fact that clinical data is often subject to various forms of noise such as missing data, inaccurate clinical records, and subjective biases in patients’ self-assessment. Existing methods generally suffer from a susceptibility to overfitting, and learning to extract task-relevant information is often suboptimal when training data is scarce. An even more glaring weakness is the sensitivity and low robustness towards noisy modality channels, where most existing models suffer from significant performance drops.

To address these issues, we propose a \underline{D}ynamic \underline{M}ultimodal \underline{I}nformation \underline{B}ottleneck (DMIB) framework, drawing inspirations from mutual information theory \cite{vera2018role,ozair2019wasserstein}, information bottleneck \cite{federici2020learning,wan2021multi,tishby2015deep}. Specifically, our DMIB consists of the following key components: i) an information bottleneck module along with dropout regularization and masking of modalities to remediate feature redundancy and model overfitting; ii) an explicit supervision to maximize the task relevant information in the final fused feature. We conduct comprehensive experiments for multimodality classification tasks across four datasets. Our method not only achieves state-of-the-art performance, but also demonstrates remarkable robustness, retaining similar performance when a modality consists of pure noise.

To summarize, our key contributions are: i) We design an information bottleneck module together with a mutual information inspired sufficiency loss which can be applied to arbitrary multimodal classification tasks as a plug-and-play module. Our fusion strategy dynamically filters out noise, maximizes the inclusion of relevant information and eliminates the need for heuristic or greedy feature selection approaches that were often employed in previous studies, leading to a superior performance. ii) To our knowledge, we are the first to conduct principled experiments to study the performance of fusion methods under different levels of noisy and redundant modalities. iii) Our method demonstrates outstanding robustness by retaining prediction performance even when noise and redundancy are introduced, making it particularly suitable for clinical settings where datasets tend to be small and biomarkers are unclear.

\begin{figure*}[h]
\begin{center}
\includegraphics[width=0.89\linewidth]
{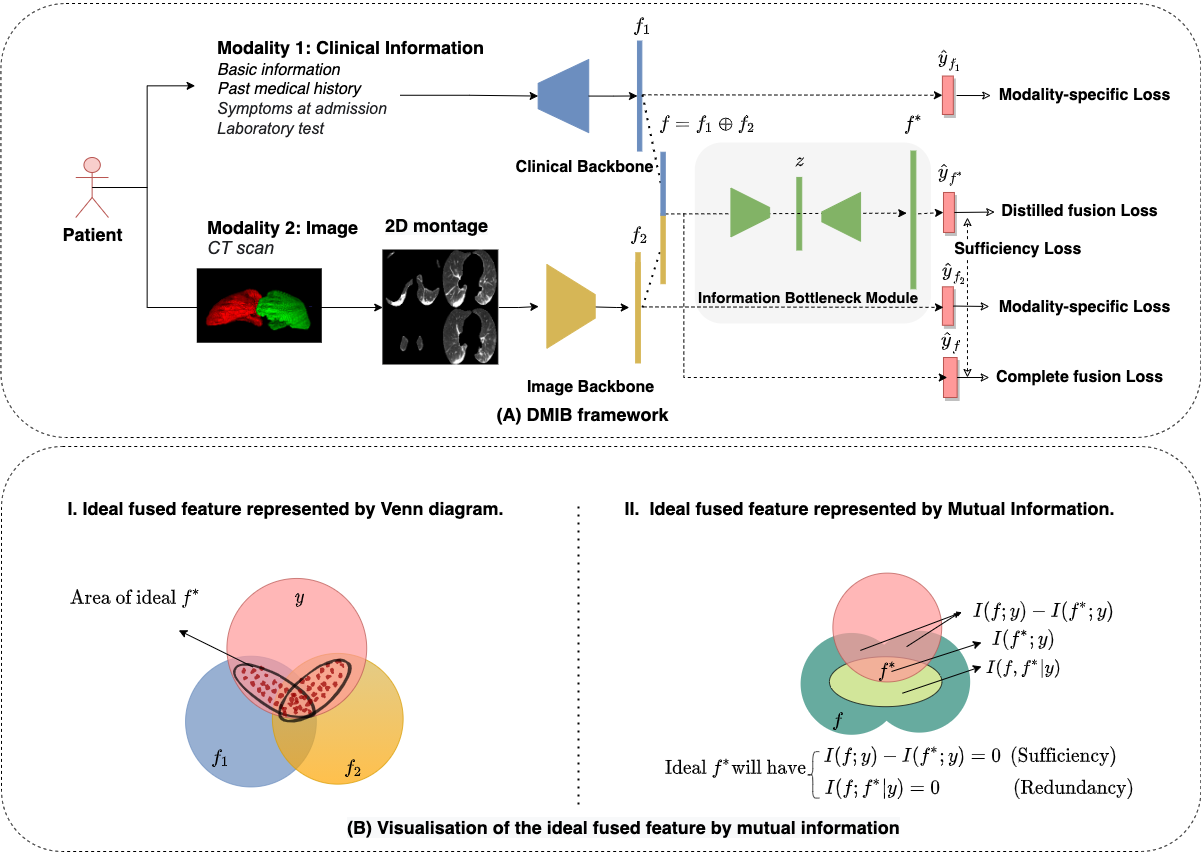}
\end{center}
\caption{(A) Illustration of our \underline{D}ynamic \underline{M}ultimodal \underline{I}nformation \underline{B}ottleneck framework for a lung disease prognosis task: a) Separate backbones are used to first extract features from each modality. b) These features are concatenated to form a preliminary fused feature $f$. c) $f$ is distilled via an information bottleneck module, arriving at a final fused feature $f^*$. d) We formulate a sufficiency loss to preserve task-relevant information in $f^*$.
(B) The mutual information between various entities can be visualized with Venn diagrams.}
\label{fig:framework}
\end{figure*}

\section{Related Works}
\subsection{Multimodal Learning for Clinical Applications}
In recent years, there has been growing interests and efforts in developing multimodal fusion for clinical applications \cite{muhammad2021comprehensive,behrad2022overview,cui2022deep,acosta2022multimodal}. 
Compared to the  early fusion cannot address the domain gaps between modalities  \cite{meng2020deep} and late fusion strategies which overlook the  cross-modality information \cite{shamout2021artificial}, intermediate fusion methods which perform fusion at the feature level \cite{polsterl2021combining,han2022multimodal} has attracted much investigation on various techniques for extracting multimodal feature learning in deep neural techniques. These methods could be broadly categorised into concatenation-based \cite{yap2018multimodal,meng2020deep,bach2017breaking}, attention-based \cite{jacenkow2022indication,duanmu2020prediction,polsterl2021combining}, projection into aligned subspaces \cite{yao2017deep,zhou2021cohesive,van2021variational,gong2022gromov}, graph-based \cite{zheng2022multi,cao2021using,keicher2021u}, and recent Transformer-based \cite{mohla2020fusatnet,xu2022multimodal,cai2022multimodal}. While concatenation methods are parameter-free and straightforward, they might not identify the intricate relationships between diverse modalities. Attention-based methods learns importance scores for intermediate multi-modality features, which allows for modeling the complex relationships between different modalities. However, introducing the attention mechanisms, along with additional network parameters increases the propensity to overfitting. Recently, a major trend in multimodal learning is to leverage transformer modules to perform cross-attention across the various modalities to obtain a fused feature representation. While transformer based multimodal learning deliver state-of-the-art performance for large scale datasets, they tend to be lackluster when training data are scarce \cite{xu2022multimodal}. 
Besides, graph models are also gaining traction in multimodal tasks with their unique ability to leverage the relations between the data from various modalities.


\subsection{Mutual Information}
Mutual information is a measure of the statistical dependency between random variables \cite{duncan1970calculation}. The core principle of deep learning approaches is to automatically learn how to extract optimal features from data, instead of manually crafting features from heuristic guidance. In this light, mutual information lies at the heart of deep learning, since it could be invoked to quantify the dependency between the learned feature representation and desired network output.

\textbf{Mutual Information Estimation}\quad \cite{linsker1988self} proposes the InfoMax principle which seeks to maximize the mutual information between feature and model output. This serves as a general prescription which is generally pertinent to neural networks and there have been many works in the recent years that explore optimal ways for mutual information estimation. One line of approach is to employ an additional neural network for estimation \cite{belghazi2018mine,hjelm2018learning} while another approach seeks variational bounds for mutual information \cite{poole2019variational}. However, mutual information remains a notoriously difficult problem due to the curse of dimensionality \cite{ozair2019wasserstein}. In other words, the amount of data samples to accurately bound mutual information scales exponentially with the dimensionality of the feature. To tackle this, researchers have proposed alternatives to mutual information, such as the Wasserstein dependency measure in \cite{ozair2019wasserstein} and the variational distillation loss in \cite{tian2021farewell}.

\textbf{Information Bottleneck}\quad Closely related to mutual information is the concept of information bottleneck. Introduced in \cite{tishby2000information}, the authors further advance this principle to put forward an explanation of the learning process in deep neural networks \cite{tishby2015deep}. Given raw data inputs, the notion of a bottleneck in the model effectuates an information distillation process, which retains only useful information, while discarding irrelevant information and superfluous noises. This is analogous to how humans learn and master knowledge and abstractions. The information bottleneck method is highly pertinent for improving model generalization and robustness and reducing overfitting \cite{vera2018role,belinkov2020variational}. This can be inherently understood from the intuition that a generalisable model should learn feature abstractions that capture the essence of the task, instead of memorizing instance-specific characteristics in the training set.

\section{Methodology}
An overview of our DMIB framework is found in Figure~\ref{fig:framework} (A). We employ separate backbones to extract features from different modalities. Supervision is applied for each backbone to guarantee the intermediate features have sufficient prediction information \footnote{To prevent the model predictions from being dominated by the modality with larger dimensions, we also enforce dimension equality by upsampling low dimensional modality features.}. Subsequently, we concatenate the extracted features for each modality to form an initial fused feature $f$, which preserves complete information from each modality. We introduce an information bottleneck module to perform information distillation, arriving at our final feature $f^*$. To ensure sufficiency of task information in $f^*$, we introduce a sufficiency loss to ensure that no task-relevant information are being discarded. 

To begin, we introduce our overall supervision objective displayed in Figure 1 (B) which motivates the design of our DMIB framework\footnote{An overview of relevant mutual information definitions is given in the Appendix.}. Subsequently, we cover the details of our information bottleneck module and our sufficiency loss.


\subsection{Overall supervision}
Given $f^*$ an encoded fused feature from the direct fused feature $f$, we would like the ideal fused feature representation to contain sufficient predictive information in $f$ while discarding all redundancy and noise. The information contained in $f^*$, is given by the mutual information between $f^*$ and $f$:
\begin{equation}
I(f; f^*)=\mathbb{E}_{p(f, f^*)}\left[\log \frac{p(f, f^*)}{p(f) p(f^*)}\right].
\end{equation}

It can be further decomposed into two components by the chain rule of mutual information \cite{federici2020learning} (given in the Appendix) as:
\begin{equation}
I(f;f^*) = I(f;f^*|y) + I(f^*;y).
\end{equation}
$I(f;f^*|y)$ quantifies the amount of task-irrelevant information in $f^*$ and $I(f^*;y)$ quantifies the predictive information in $f^*$ for task $y$. The goal is to simultaneously minimize redundancy $I(f;f^*|y)$ and maximize task relevance $I(f^*;y)$, which can be combined into the information bottleneck objective \cite{tishby2000information}:
\begin{equation}
\begin{aligned}\label{lagrangian}
\min_{{f^*}}\mathcal{L}_{IB}=I(f;f^*|y) - \gamma I(f^*;y)\end{aligned}.
\end{equation}
However, estimating for mutual information in high dimensions is intractable in general \cite{poole2019variational,nguyen2010estimating} due to the curse of dimensionality \cite{ozair2019wasserstein}, \emph{i.e.} accurate estimation of the expected information in empirical distributions requires a sample size that scales exponentially with the dimension of the data. In light of this, we do not explicitly optimize for Eqn. \eqref{lagrangian} via mutual information estimation. Instead, we delegate information distillation to our information bottleneck module in Section \ref{sec:ib}. For maximizing the task-relevant information $I(f^*;y)$, we formulate a tractable sufficiency loss in Section \ref{sec:loss}. The joint framework consisting of the information bottleneck module and the sufficiency loss enables $f^*$ to converge to an ideal feature with maximal relevance and minimal redundancy.

\subsection{Information Bottleneck Module} \label{sec:ib}
Given $n$ features $f_1, \cdots, f_n$ extracted from $n$ modalities, we first generate an initial fused feature $f$ by direct concatenation with a masking
$$f = \bigoplus_{i=1}^n m_i f_i,$$
where $f \in \mathbb{R}^N$ and $m_i$ denotes a masking coefficient. Specifically, at each training iteration, we uniformly sample a random number $u\sim U([0,1])$, and the masking coefficients are formally given by
$m_i = \mathbbm{1}_{[\frac{i-1}{2n)};\frac{i}{2n}]}(u).$
In other words, the model can only access the full features across all modalities for half of the training iterations, while for the other half, a single modality $f_i$ is masked out. The motivation of masking individual modalities is to allow the model to function in the absence of data-streams from a modality, thus improving robustness \cite{he2022masked}. 
Subsequently, our information bottleneck module comprises of two linear projection layers along with dropouts and ReLU. The operations may be summarized as:
\begin{equation}
f \xrightarrow[\text{Dropouts + ReLU}]{\text{Linear Projection}} z \xrightarrow[\text{Dropouts + ReLU}]{\text{Linear Projection}} f^*.
\end{equation}
$z$ is of dimension $p<n$ while the final feature representation $f^*$ has $n$ dimensions. The purpose of re-projecting $z$ back to the same dimensions as $f$ is in order to perform more effective feature-level supervision of the distilled feature and the initial feature to learn a more predictive feature $f^*$, which is introduced in the next section. It is observed that the feature-level supervision is more effective when $f^*$ and the initial feature $f$ is aligned with equal dimensions.

\subsection{Sufficiency Loss} \label{sec:loss}
In this section, we introduce our sufficiency loss which serves as a feature-level supervision to maximize the task-relevant information $I(f^*;y)$ in Eqn.~\eqref{lagrangian}. Since $f^*$ is a feature extracted from $f$, the information contained in $f^*$ cannot exceed that of $f$, and $I(f^*;y)\le I(f;y)$. Maximizing $I(f^*;y)$ is therefore equivalent to: 
\begin{equation}\label{sufficiency}
\min I(f ; y)-I(f^*;y).
\end{equation}
To solve this optimization problem, we make use of the following proposition \cite{tian2021farewell,liu2022temporal} 
(proof can be found in the Appendix):
\begin{equation}
KL[ p(y|f) \| p(y|f^*) ] = 0 \implies I(y;f) - I(y;f^*) = 0,
\end{equation}
where $KL$ denotes the Kullback-Leibler divergence.
Therefore, instead of having to estimate the mutual information in Eqn. \eqref{sufficiency}, we formulate our sufficiency loss as\begin{equation} \label{eqn:VSD}
\L_{\text{Sufficiency}}= KL[p_{\theta}(y|f) \| p(y|f^*)].
\end{equation}

\subsection{Overall objective}
The overall loss function in our DMIB framework is:
\begin{eqnarray}\label{overall loss}
\L = \L_{f}(y, \hat{y}_f) + 
\alpha \sum_{i} \L_{\text{modality}}(y, \hat{y}_{f_i}) + \nonumber\\ \underbrace{ 
\L_{f^*}(y,\hat{y}_{f^*}) + \beta \L_{\text{sufficiency}}(\hat{y}_{f}, \hat{y}_{f^*})}_{\text{Supervision of Information Bottleneck Module}}
\end{eqnarray}
where $\L_f$ and $\L_{f^*}$ are respective losses to supervise the direct fused expression datafeature and distilled fused features, and $\L_{\text{modality}}$ supervises each modality backbone.
Here, $\alpha$ and $\beta$ are model hyperparameters to control the roles of modal-specific supervision and feature-level supervision, respectively. $\hat{y}_{f}$ and $\hat{y}_{f^*}$ are the classifier results from the observed fused feature and distilled fused feature.

\section{Experiments}
\subsection{Datasets}
We trained our proposed DMIB for various multimodal medical classification tasks on four datasets: 
i) Our in-house \textbf{ITAC dataset}, which includes HRCT and 18 clinical features of 566 COVID19 inpatients, is utilized for predicting the 10-day mortality rate of COVID19 inpatients; 
ii) the public \textbf{iCTCF dataset}  \cite{ning2020open}\footnote{Available at \url{https://ngdc.cncb.ac.cn/ictcf/}}, which includes HRCT scans and up to 81 clinical features, is used to predict the morbidity outcomes of 751 COVID19 patients and the detection of 751 COVID19 patients from 529 non-COVID19 patients respectively;
iii) \textbf{BRCA dataset} for diagnosis of breast carcinoma PAM50 subtypes (mRNA expression data, DNA methylation data, and miRNA expression data) of 875 patients \cite{wang2020moronet}; 
iv) \textbf{ROSMAP dataset} for diagnosis of Alzheimer's Disease from 351 patients \cite{wang2020moronet} \footnote{iii) and iv) are available at \url{https://github.com/txWang/MOGONET}}. 
A summary of the various datasets is found in Table~\ref{tab:datasets}. It is worth mentioning that ITAC and iCTCF are collected from different countries and have no overlap in enrolled patients.

\begin{table*}[h!]
\centering
\caption{Summary of datasets}\label{tab:datasets}
\resizebox{1\textwidth}{!}
{
\begin{tabular}{|l|l|l|l|l|}
\hline
Dataset & Modality & Task Description & Enrolled Patients & Generated montages \\ \hline
ITAC & HRCT scans + up to 18 clinical features & Prognosis for COVID19 mortality in 10 Days & Deceased: 257 / Cured: 309 & Deceased: 3084 / Cured: 3090  \\ \hline
\multirow{2}{*}{iCTCF} & \multirow{2}{*}{HRCT scans + up to 81 clinical features} & Prognosis for COVID19 morbidity outcome & Severe symptoms: 202 / Mild symptoms: 549 & Severe symptoms: 606 / Mild symptoms: 549 \\ \cline{3-5}

& & Diagnosis for COVID19 patients & PCR positive: 751 / PCR negative: 529 & PCR positive: 3755 / PCR negative: 3174 \\ \hline

BRCA & mRNA, DNA methylation, miRNA & Diagnosis for breast carcinoma PAM50 subtype & Normal: 115 / Basal: 131 / Her2: 46 / LumA: 436 / LumB: 147 & / \\ \hline
ROSMAP & mRNA, DNA methylation, miRNA & Diagnosis for Alzheimer's Disease & Normal: 169 / AD: 182 & / \\ \hline
\end{tabular}
}
\end{table*}

\begin{table*}[h]
\caption{Performance of various methods on the ITAC dataset. Bold denotes the clinical settings where each methods achieves best AUC.}
\label{tab:ITAC}
\centering
\resizebox{\textwidth}{!}{
\begin{tabular}{c|cccccc|c|cccccc}
\toprule[1.5pt]
& \multicolumn{6}{c|}{\textbf{Test}} & &\multicolumn{6}{c}{\textbf{Test}} \\ \hline
& AUC & Accuracy & Sensitivity & Specificity & $\mbox{JW}_{0.5}$ & $\mbox{JW}_{0.6}$ & & AUC & Accuracy & Sensitivity & Specificity & $\mbox{JW}_{0.5}$ & $\mbox{JW}_{0.6}$ \\ \hline
& \multicolumn{6}{c|}{Single modality} & & \multicolumn{6}{c}{Fusion: Concatenation \cite{meng2020deep}} \\
\hline
clinical 1 & 0.76 & 0.69 & 0.75 & 0.63 & 0.69 & 0.70& clinical 1 & 0.77 & 0.71 & 0.74 & 0.68 & 0.71& 0.72 \\
\textbf{clinical 4} & \textbf{0.82} & 0.75 & 0.77 & 0.73 &0.75&0.75 & clinical 4 & 0.78 & 0.69 & 0.65 & 0.72 &0.69 &0.68 \\
\textbf{clinical 7} & \textbf{0.82} & 0.75 & 0.74 & 0.76 &0.75& 0.75 &\textbf{ clinical 7} & \textbf{0.83} & 0.76 & 0.72 & 0.78 & 0.75 &0.74 \\
clinical 18 & 0.81 & 0.73 & 0.71 & 0.76 &0.74& 0.73 & clinical 18 & 0.80 & 0.74 & 0.62 & 0.85 &0.74 &0.71 \ \\
CT only & 0.77 & 0.72 & 0.57 & 0.85 &0.71&0.68 & pure noise & 0.74 & 0.66 & 0.72 & 0.62 &0.67 &0.68 \\
\hline
& \multicolumn{6}{c|}{Fusion: Attention \cite{duanmu2020prediction}}
& 
& \multicolumn{6}{c}{Fusion: Transformer \cite{mohla2020fusatnet}} \\
\hline
\textbf{clinical 1} & \textbf{0.81} & 0.76 & 0.72 & 0.78 & 0.75 & 0.74 & clinical 1 & 0.66 & 0.65 & 0.54 & 0.74 & 0.64& 0.62 \\
\textbf{clinical 4} & \textbf{0.81} & 0.74 & 0.77 & 0.72 & 0.75 & 0.75 & clinical 4 & 0.64 & 0.57 & 0.51 & 0.62 & 0.62 & 0.55 \\
clinical 7 & 0.68 & 0.62 & 0.51 & 0.71 & 0.61 & 0.59 & \textbf{clinical 7} & \textbf{0.73} & 0.67 & 0.66 & 0.68 & 0.67 &0.67\\
clinical 18 & 0.76 & 0.67 & 0.65 & 0.69 & 0.67 & 0.66 & \textbf{clinical 18} & \textbf{0.73} & 0.66 & 0.65 & 0.67 & 0.66 & 0.66 \\ 
pure noise & 0.65 & 0.62 & 0.51 & 0.72 & 0.61 & 0.59 & pure noise & 0.65 & 0.64 & 0.54 & 0.73 & 0.64 &0.62\\
\hline

& \multicolumn{6}{c|}{Fusion: Dynamic \cite{han2022multimodal}}
& 
& \multicolumn{6}{c}{Fusion: Graph \cite{zheng2022multi}} \\
\hline
clinical 1 & 0.82 & 0.76 & 0.58 & 0.91 & 0.75	& 0.71 & clinical 1 & 0.73	&  0.72 & 0.67 & 0.77 & 0.72 & 	0.71   \\
clinical 4 & 0.83 & 0.70 & 0.85 & 0.58 &  0.72&0.74 & clinical 4 & 0.69	&0.68&	0.62	&0.74&	0.68&	0.67    \\
\textbf{clinical 7} &  \textbf{0.85} & 0.76 & 0.77 & 0.74 & 0.76 & 0.76  & \textbf{clinical 7} & \textbf{0.76} &	0.75&	0.75&	0.78&	0.77&	0.76 \\
clinical 18 & 0.83 & 0.74 &  0.88 & 0.63  & 0.75 & 0.78 & clinical 18 & 0.75	&0.75&	0.74&	0.76	&0.75	&0.75 \\ 
pure noise & 0.75 & 0.66 &  0.66 & 0.67 & 0.67 & 0.66 & pure noise &0.68&	0.67	&0.60&	0.74	&0.67	&0.66 \\
\hline

& \multicolumn{6}{c|}{Fusion: Subspace \cite{zhou2021cohesive}}
& 
& \multicolumn{6}{c}{Fusion: DMIB}\\
\hline
\textbf{clinical 1} & \textbf{0.84} & 0.75 & 0.69 & 0.79 & 0.74 & 0.73 & clinical 1 & 0.85 & 0.76 & 0.82 & 0.72 & 0.77 & 0.78 \\
\textbf{clinical 4} & \textbf{0.84} & 0.76 & 0.75 & 0.76 &0.76 & 0.75 & clinical 4 & 0.85 & 0.78 & 0.78 & 0.77 & 0.78 &0.78 \\
\textbf{clinical 7} & \textbf{0.84} & 0.78 & 0.77 & 0.78 & 0.78 & 0.77 & \textbf{clinical 7} & \textbf{0.86} & 0.80 & 0.75 & 0.83 & 0.79 &0.78 \\
clinical 18 & 0.83 & 0.76 & 0.57 & 0.92 & 0.75 & 0.71 & \textbf{clinical 18} & \textbf{0.86} & 0.76 & 0.77 & 0.76 & 0.77 & 0.77 \\
pure noise & 0.71 & 0.63 & 0.55 & 0.69 & 0.62 & 0.61 & pure noise & 0.76 & 0.72 & 0.66 & 0.77 & 0.72 & 0.70 \\
\bottomrule[1.5pt]
\end{tabular}}
\end{table*}

\begin{table*}[h]
\caption{Performance of various methods on the iCTCF dataset for morbidity prediction. Bold denotes the best AUC in each fusion setting.}
\label{tab:iCTCF}
\centering
\resizebox{\textwidth}{!}{
\begin{tabular}{c|cccccc|c|cccccc}
\toprule[1.5pt]
& \multicolumn{6}{c|}{\textbf{Test}} & &\multicolumn{6}{c}{\textbf{Test}} \\ \hline
& AUC & Accuracy & Sensitivity & Specificity & $\mbox{JW}_{0.5}$ & $\mbox{JW}_{0.6}$ & & AUC & Accuracy & Sensitivity & Specificity & $\mbox{JW}_{0.5}$ & $\mbox{JW}_{0.6}$ \\ \hline
& \multicolumn{6}{c|}{Single modality} & & \multicolumn{6}{c}{Fusion: Concatenation \cite{meng2020deep}} \\
\hline
clinical 1 & 0.72 & 0.62 & 0.71 & 0.59 & 0.65 & 0.66 & clinical 1 & 0.71 & 0.63 & 0.67 & 0.62 & 0.64 & 0.65 \\
All clinical & 0.74 & 0.69 & 0.67 & 0.70 & 0.68 & 0.68 & All clinical & 0.77 & 0.72 & 0.65 & 0.75 & 0.70 & 0.69 \\
CT only  & 0.71 & 0.68 & 0.63 & 0.70 & 0.67 & 0.66 & \textbf{pure noise}  & \textbf{0.71} & 0.65 & 0.67 & 0.64 & 0.66 & 0.66  \\
\hline
& \multicolumn{6}{c|}{Fusion: Attention \cite{duanmu2020prediction}}
& 
& \multicolumn{6}{c}{Fusion: Transformer \cite{mohla2020fusatnet}} \\
\hline
clinical 1 &  0.72 & 0.65 & 0.63 & 0.65 & 0.64 & 0.64  & clinical 1 & 0.59 & 0.74 & 0.06 & 0.99 & 0.53 & 0.43 \\
All clinical & 0.77 & 0.75 & 0.65 & 0.79 & 0.72 & 0.70  & All clinical & 0.67 &  0.67 & 
 0.59 &  0.70 &  0.65 &  0.63   \\ 
pure noise  & 0.56 & 0.69 & 0.31 & 0.83 & 0.57 & 0.52 & pure noise & 0.65 & 0.38 & 0.90 & 0.18 & 0.54 & 0.61 \\
\hline
& \multicolumn{6}{c|}{Fusion: Dynamic \cite{han2022multimodal}}& & \multicolumn{6}{c}{Fusion: Graph \cite{zheng2022multi}} \\
\hline
clinical 1 & 0.78 & 0.75 & 0.16 & 0.96 & 0.56 & 0.48 & clinical 1 & 0.71 & 0.70 & 0.67 & 0.72 & 0.70 & 0.69 \\
All clinical & 0.78 & 0.50 & 0.90 & 0.35 & 0.62 & 0.68 & All clinical & 0.74 & 0.72 & 0.67 & 0.77 & 0.72 & 0.71 \\ 
pure noise  & 0.67 & 0.73 & 0.00 & 1.00 & 0.50 & 0.40 & pure noise & 0.62 & 0.64 & 0.69 & 0.60 & 0.65 & 0.65\\
\hline

& \multicolumn{6}{c|}{Fusion: Subspace \cite{zhou2021cohesive}}& & \multicolumn{6}{c}{Fusion: DMIB}\\ \hline
clinical 1 & 0.78 & 0.63& 0.88&0.54&0.71&0.75 & \textbf{clinical 1} & \textbf{0.79} & 0.70 & 0.76 & 0.68 & 0.72 & 0.73  \\
All clinical & 0.80 & 0.68 & 0.76 & 0.64 & 0.70 & 0.72 & \textbf{All clinical} & \textbf{0.82} & 0.73 & 0.71 & 0.74 & 0.72 & 0.72\\ 
\textbf{pure noise}  & \textbf{0.71} & 0.58 & 0.75 & 0.52 & 0.63 & 0.66  & \textbf{pure noise} & \textbf{0.71} & 0.63 & 0.69 & 0.62 & 0.65 & 0.66 \\
\bottomrule[1.5pt]
\end{tabular}}
\end{table*}

\begin{table*}[h]
\caption{Performance of various methods on the iCTCF dataset for COVID19 diagnosis. Bold denotes the best AUC in each fusion setting.}
\label{tab:iCTCFdiagnosis}
\centering
\resizebox{\textwidth}{!}{
\begin{tabular}{c|cccccc|c|cccccc}
\toprule[1.5pt]
& \multicolumn{6}{c|}{\textbf{Test}} & &\multicolumn{6}{c}{\textbf{Test}} \\ \hline
& AUC & Accuracy & Sensitivity & Specificity & $\mbox{JW}_{0.5}$ & $\mbox{JW}_{0.6}$ & & AUC & Accuracy & Sensitivity & Specificity & $\mbox{JW}_{0.5}$ & $\mbox{JW}_{0.6}$ \\ \hline
& \multicolumn{6}{c|}{Single modality} & & \multicolumn{6}{c}{Fusion: Concatenation \cite{meng2020deep}} \\
\hline
clinical 1  & 0.51 &  0.51 &  0.55 &  0.46 &  0.51 &  0.52 &  \textbf{clinical 1} &\textbf{0.81}&0.74&0.72&0.76&0.74&0.73 \\
All clinical & 0.80 & 0.74 & 0.77 & 0.70 & 0.73 & 0.74 & All clinical & 0.85&0.78&0.85&0.68&0.76&0.78 \\
CT only  & 0.80&0.72&0.85&0.55&0.70&0.73 & \textbf{pure noise} & \textbf{0.81} &0.73&0.81&0.61&0.71&0.73 \\
\hline
& \multicolumn{6}{c|}{Fusion: Attention \cite{duanmu2020prediction}} &  & \multicolumn{6}{c}{Fusion: Transformer \cite{mohla2020fusatnet}} \\
\hline
clinical 1 & 0.70 & 0.67 & 0.74 & 0.58 & 0.66 & 0.68 & clinical 1 & 0.62 & 0.63 & 0.86 & 0.29 & 0.58 & 0.63  \\
All clinical & 0.80&0.75&0.80&0.69&0.74&0.76  & All clinical & 0.66& 0.64 & 0.68 &  0.59 & 0.63  & 0.64    \\ 
pure noise & 0.50&0.49&0.61&0.33&0.47&0.50 & pure noise &0.64&0.59&0.60&0.59&0.59&0.59   \\
\hline
& \multicolumn{6}{c|}{Fusion: Dynamic \cite{han2022multimodal}}
& 
& \multicolumn{6}{c}{Fusion: Graph \cite{zheng2022multi}} \\
\hline
clinical 1 &  0.80 & 0.74 & 0.85 & 0.59 & 0.72 & 0.74 & clinical 1 & 0.74 & 0.73 & 0.74 &  0.73 &  0.74 & 0.74 \\
All clinical &0.86&0.76&0.86&0.62&0.74&0.76 &All clinical & 0.83& 0.83 &  0.84 & 0.83  &  0.84 &   0.84 \\ 
pure noise & 0.80&0.74&0.82&0.62&0.72&0.74 & pure noise& 0.73 & 0.73 & 0.72&0.74 &0.73  &0.73   \\
\hline

& \multicolumn{6}{c|}{Fusion: Subspace \cite{zhou2021cohesive}}& & \multicolumn{6}{c}{Fusion: DMIB}\\\hline 
clinical 1 & 0.79 & 0.70 & 0.80 & 0.56 & 0.68 & 0.71 & clinical 1 & 0.80 & 0.73 & 0.86 & 0.53 & 0.70 & 0.73   \\
All clinical &0.86&0.77&0.79&0.74&0.77&0.77& \textbf{All clinical} &\textbf{0.89}&0.82&0.78&0.87&0.83&0.82\\ 
pure noise & 0.80&0.72&0.95&0.40&0.67&0.73 & pure noise & 0.80&0.73&0.74&0.71&0.73&0.73 \\ 
\bottomrule[1.5pt]
\end{tabular}}
\end{table*}

\subsection{Implementation details}
\noindent \textbf{ITAC and iCTCF.} For data preprocessing, we crop the lung regions from the complete 3D scan and resize each slice to the dimensions of 350 $\times$ 350 pixels, preserving its original depth. Subsequently, we generate  2D montages for each patient from their 3D HRCT scans. Each montage is composed of 4 randomly chosen axial slices, with each slice originating from one of the 4 equally divided regions of the scan, placed on a 2x2 grid. The utilization of 2D montages enables us to generate diverse montages for each scan without duplication. This approach can effectively mitigate the challenges of data scarcity and data imbalance commonly encountered in medical image datasets. It offers an advantage over using the entire scan as input, as well as compared to single-slice input, by preserving an adequate amount of predictive information in the input. The number of generated montages for each class is summarized in Table ~\ref{tab:datasets}.

For training and testing the proposed model, we conduct patient-level data splitting to prevent information leakage. We reserve 20\% of patients from each class for testing and using the remaining patients for model training with five-fold cross-validation. Patients are divided into five folds with no overlaps using stratified sampling, with each having the same patient distribution. Each of this fold serves as a validation set, and we train five models from scratch on the remaining four folds. From the five models, the model attaining the top AUC on its validation set is chosen as the final testing model. For testing, we generate 10 montages for each patient and take the median prediction over the 10 montages as the final result.

We employ DensetNet-121 \cite{huang2017densely} as the backbone for the CT montage modality to obtain an image feature $f_1\in\mathbb{R}^{1024}$. For the clinical data, we perform mean imputation for missing records followed by data normalization. We employ four linear layers to obtain a clinical feature $f_2\in\mathbb{R}^{1024}$. The dimension of $z$ in the IB module is set to 1024 and loss function weights are set to $\alpha=1, \beta=10$. We use the cross-entropy loss for $\L_f, \L_{f^*}, \L_\text{modality}$. All training are done with the Adam optimizer (initial learning rate of 1e-6 and a linear decay rate of 1e-2), a batch size of 8, and trained for 70 epochs on a single RTX3090.

\noindent \textbf{ROSMAP and BRCA.} For fair comparison, we adopt similar experimental settings as \cite{han2022multimodal}, and we reenacted their experiments using their open-sourced codes. We set the dimension of $z$ to 1024 for the bottleneck.

\subsection{Evaluation setting}
\noindent \textbf{ITAC.} To evaluate the robustness of different fusion strategies in handling noises and redundant information, we consider multiple experimental settings, starting with the most informative and complete clinical variables, and gradually expanding to those with larger degrees of redundancy and noise. Specifically, we fix five settings: i) 1 clinical variable (age) which is the most informative feature. ii) 4 clinical variables (i) + oxygen saturation, platelets, measured saturation oxygen, which are also deemed as relevant by domain experts. iii) 7 clinical variables (ii) + respiratory rate, pO2, and D-Dimer, which are informative but have a considerable 30\% of the data are missing. iv) (iii) + reported symptoms and health records, which are more prone to missing data and subjective biases in the symptoms description, and deemed to be of secondary importance. v) 1 random integer between 0 and 100 that serves as a noise input replacing age. In the Appendix, we further include the details of the clinical data for our ITAC dataset.

We benchmark DMIB against single modality models (CT backbone and clinical backbone in our method), as well as various multimodal fusion schemes including direct concatenation (Concatenation \cite{meng2020deep}), fusion via channelwise attention (Attention \cite{duanmu2020prediction}), fusion via transformer-based cross-attention  (Transformer \cite{mohla2020fusatnet}), fusion via dynamic weighing of each modality channel (Dynamic \cite{han2022multimodal}),  fusion via projection to a common subspace (Fusion: Subspace \cite{zhou2021cohesive}) and graph-based fusion (Graph \cite{zheng2022multi}). We adopt the Area Under Receiver Operating Characteristic Curve (AUC) as the primary evaluation metric, with accuracy, sensitivity, specificity, and the weighted Youden indices $JW_{0.5}$ and $JW_{0.6}$ as auxiliary metrics. Results are reported in Table~\ref{tab:ITAC}.

\noindent \textbf{iCTCF.} To eliminate the manual selection of optimal clinical data combinations for improved fusion performance, we extend our evaluation on this dataset to assess the robustness of the proposed model for both diagnostic and prognostic tasks. We integrate the CT scans with all 81 available clinical variables, to showcase the superiority of DMIB in fusing potentially redundant and noisy data without requiring manual selection. Furthermore, we perform fusion with single variables, `Age' and `pure noise' (consisting of random age and temperature), as reference points for assessing fusion performance with all clinical information.

\noindent \textbf{BRCA \& ROSMAP.} To further evaluate the generalization of DMIB to other modalities, we also performed experiments on the  BRCA and ROSMAP dataset consisting of multimodal genomic data. We benchmark against Fusion: Dynamic \cite{han2022multimodal}, the state-of-the-art method which employs fusion strategy consisting of model-specific supervision, attention, sparsity constraints and dynamic assignment of confidence to each modality. Following \cite{han2022multimodal}, we report AUC, WeightedF1 and MAcroF1 and F1 in Table \ref{tab:BRCAROSMAP}.

\begin{table}[h]
\centering
\caption{Performance of multimodal methods on BRCA and ROSMAP. $^{\dagger}$ denotes the result showing significant difference to that of DMIB with $p < 1e-4$ by paired T test.}
\label{tab:BRCAROSMAP}
\resizebox{1\linewidth}{!}{
\begin{tabular}{c|ccc|ccc}
\toprule[1.5pt]
& \multicolumn{3}{c|}{BRCA} & \multicolumn{3}{c}{ROSMAP} \\ \hline
Method & \multicolumn{1}{c|}{{ACC}} & \multicolumn{1}{c|}{WeightedF1} & MacroF1 & \multicolumn{1}{c|}{ACC} & \multicolumn{1}{c|}{F1} & AUC$^\dagger$ \\ \hline
Fusion: Dynamic & \multicolumn{1}{c|}{\textbf{87.1+0.5$^\dagger$}} & \multicolumn{1}{c|}{\textbf{87.4+0.6$^\dagger$}} & \textbf{83.5+0.9$^\dagger$} & \multicolumn{1}{c|}{81.7+1.5$^\dagger$} & \multicolumn{1}{c|}{82.3+1.5$^\dagger$} & 90.0+1.2$^\dagger$ \\ \hline
Fusion: DMIB & \multicolumn{1}{c|}{86.0+0.7} & \multicolumn{1}{c|}{86.0+0.8} & 81.6+0.9 & \multicolumn{1}{c|}{\textbf{84.9+1.8}} & \multicolumn{1}{c|}{{\textbf{85.3+1.7}}} & {\textbf{91.6+0.7}} \\
\bottomrule[2pt]
\end{tabular}
}
\end{table}

\begin{table*}[h]
\caption{Ablation studies on ITAC and BRCA datasets}
\label{tab:ablation}
\centering 
\resizebox{\textwidth}{!}{
\begin{tabular}{lllllllllllll}
\toprule[1.5pt] 
\multicolumn{1}{c|}{} & \multicolumn{5}{c|}{Ablation Setting} & \multicolumn{4}{c|}{ITAC} & \multicolumn{3}{c}{BRCA} \\ \hline

\multicolumn{1}{c|}{No.} & \multicolumn{1}{c}{$\L_f$} & \multicolumn{1}{c}{$\mbox{IB}$} & \multicolumn{1}{c}{$\L_{f^*}$} & \multicolumn{1}{c}{$\L_{\mbox{sufficiency}}$} & \multicolumn{1}{c|}{$\L_{\mbox{modality}}$} & \multicolumn{1}{c|}{AUC} & \multicolumn{1}{c|}{Accuracy} & \multicolumn{1}{c|}{Sensitivity} & \multicolumn{1}{c|}{Specificity} & \multicolumn{1}{c|}{ACC} & \multicolumn{1}{c|}{WeightedF1} & \multicolumn{1}{c}{MacroF1} \\ \hline

\multicolumn{1}{c|}{1.} & \multicolumn{1}{c}{\check} & \multicolumn{1}{c}{\blank} & \multicolumn{1}{c}{\blank} & \multicolumn{1}{c}{\blank} & \multicolumn{1}{c|}{\blank} & \multicolumn{1}{c|}{0.83} & \multicolumn{1}{c|}{0.78} & \multicolumn{1}{c|}{0.65} & \multicolumn{1}{c|}{0.90} & \multicolumn{1}{c|}{81.4+0.6$^\dagger$} & \multicolumn{1}{c|}{81.1+0.7$^\dagger$} & \multicolumn{1}{c}{76.2+1.0$^\dagger$} \\ \cline{7-13}

\multicolumn{1}{c|}{2.} & \multicolumn{1}{c}{\blank} & \multicolumn{1}{c}{\check} & \multicolumn{1}{c}{\check} & \multicolumn{1}{c}{\blank} & \multicolumn{1}{c|}{\check} & \multicolumn{1}{c|}{0.83} & \multicolumn{1}{c|}{0.77} & \multicolumn{1}{c|}{0.74} & \multicolumn{1}{c|}{0.79} & \multicolumn{1}{c|}{81.0+1.0$^\dagger$} & \multicolumn{1}{c|}{80.2+1.2$^\dagger$} & \multicolumn{1}{c}{75.3+1.2$^\dagger$} \\ \cline{7-13}

\multicolumn{1}{c|}{3.} & \multicolumn{1}{c}{\check} & \multicolumn{1}{c}{\check} & \multicolumn{1}{c}{\check} & \multicolumn{1}{c}{\blank} & \multicolumn{1}{c|}{\check} & \multicolumn{1}{c|}{0.83} & \multicolumn{1}{c|}{0.75} & \multicolumn{1}{c|}{0.69} & \multicolumn{1}{c|}{0.79} & \multicolumn{1}{c|}{81.1+0.8$^\dagger$} & \multicolumn{1}{c|}{80.4+0.8$^\dagger$} & \multicolumn{1}{c}{75.4+1.0$^\dagger$}  \\ \cline{7-13}

\multicolumn{1}{c|}{4.} & \multicolumn{1}{c}{\check} & \multicolumn{1}{c}{\check} & \multicolumn{1}{c}{\check} & \multicolumn{1}{c}{\check} & \multicolumn{1}{c|}{\blank} & \multicolumn{1}{c|}{0.85} & \multicolumn{1}{c|}{0.77} & \multicolumn{1}{c|}{0.75} & \multicolumn{1}{c|}{0.78} & \multicolumn{1}{c|}{83.9+1.5$^\dagger$} & \multicolumn{1}{c|}{83.8+1.7$^\dagger$} & \multicolumn{1}{c}{79.7+1.8$^\dagger$} \\ \cline{7-13}

\multicolumn{1}{c|}{5.} & \multicolumn{1}{c}{\check}& \multicolumn{1}{c}{\check} & \multicolumn{1}{c}{\blank} & \multicolumn{1}{c}{\check} & \multicolumn{1}{c|}{\check} & \multicolumn{1}{c|}{0.84} & \multicolumn{1}{c|}{0.78} & \multicolumn{1}{c|}{0.65} & \multicolumn{1}{c|}{0.88} & \multicolumn{1}{c|}{86.1+0.5$^\dagger$} & \multicolumn{1}{c|}{86.2+0.6$^\dagger$} & \multicolumn{1}{c}{81.8+0.7$^\dagger$} \\ \cline{7-13}

\multicolumn{1}{c|}{6.} & \multicolumn{1}{c}{\check} & \multicolumn{1}{c}{\check} & \multicolumn{1}{c}{\check} & \multicolumn{1}{c}{\check} & \multicolumn{1}{c|}{\check} & \multicolumn{1}{c|}{0.86} & \multicolumn{1}{c|}{0.76} & \multicolumn{1}{c|}{0.77} & \multicolumn{1}{c|}{0.76} & \multicolumn{1}{c|}{86.0+0.7$^\dagger$} & \multicolumn{1}{c|}{86.0+0.8$^\dagger$} & \multicolumn{1}{c}{81.6+0.9$^\dagger$} \\
\bottomrule[1.5pt]
\end{tabular}
}
\end{table*}

\begin{table}[h]
\centering
\caption{Experimenting with different image backbones and unchanged clinical backbone on ITAC (with all clinical variables). Bond fonts denote the best result for each backbone.}
\label{tab:backbones}
\resizebox{\linewidth}{!}{
\begin{tabular}{c|c|cccccc}
\toprule[1.5pt]
Image Backbone & Experiment & AUC & ACC & Sens & Spec & JW5 & JW6 \\ \hline
 & CT Only & 0.77 & 0.71 & 0.68 & 0.74 & 0.71 & 0.7 \\ \cline{2-8} 
ResNet18 \cite{he2016deep} & Concat & 0.77 & 0.72 & 0.7 & 0.75 & 0.73 & 0.72 \\ \cline{2-8} 
 & Subspace & 0.79 & 0.73 & 0.73 & 0.73 & 0.73 & 0.73 \\ \cline{2-8}
 & \textbf{Proposed} & \textbf{0.83} & \textbf{0.77} & \textbf{0.77} & \textbf{0.77} & \textbf{0.77} & \textbf{0.77} \\ \hline
 & CT Only & 0.71 & 0.69 & 0.75 & 0.63 & 0.69 & 0.7 \\ \cline{2-8} 
ResNet50 \cite{he2016deep} & Concat & 0.75 & 0.67 & 0.74 & 0.62 & 0.68 & 0.69 \\ \cline{2-8} 
 & Subspace & 0.77 & 0.72 & 0.75 & 0.69 & \textbf{0.72} & \textbf{0.73} \\ \cline{2-8}
 & \textbf{Proposed} & \textbf{0.81} & \textbf{0.72} & \textbf{0.74} & \textbf{0.71} & \textbf{0.72} & \textbf{0.73} \\ \hline
 & CT Only & 0.71 & 0.62 & 0.58 & 0.65 & 0.62 & 0.61 \\ \cline{2-8} 
EfficientNet\_b0 \cite{tan2019efficientnet} & Concat & 0.76 & 0.66 & 0.72 & 0.62 & 0.67 & 0.68 \\ \cline{2-8} 
 & Subspace & 0.74 & 0.65 & 0.71 & 0.60 & 0.66 & 0.67 \\ \cline{2-8}
 &\textbf{Proposed} & \textbf{0.82} & \textbf{0.73} & \textbf{0.74} & \textbf{0.72} & \textbf{0.73} & \textbf{0.73} \\ \hline 
 & CT Only & 0.69 & 0.62 & 0.62 & 0.63 & 0.62 & 0.62 \\ \cline{2-8} 
PoolFormer\_v2\_tiny \cite{yu2022metaformer} & Concat & 0.73 & 0.67 & 0.66 & 0.68 & 0.67 & 0.67 \\ \cline{2-8} 
 & Subspace & 0.77 & 0.71 & 0.71 & 0.72 & 0.71 & 0.71 \\ \cline{2-8}
 & \textbf{Proposed} & \textbf{0.79} & \textbf{0.73} & \textbf{0.74} & \textbf{0.72} & \textbf{0.73} & \textbf{0.73} \\
\bottomrule[1.5pt]
\end{tabular}
}
\end{table}

\subsection{Results}
\noindent \textbf{ITAC.} 
As reported in Table~\ref{tab:ITAC}, our DMIB outperforms all competing methods across all settings for our ITAC dataset. As expected, we observe that inclusion of more clinical variables often fails to improve prognostic performance in many existing methods. In fact, the inclusion of more modalities in attention-based, transformer-based and graph-based fusion might even lead to weaker performance than single modalities. Such fluctuations across different experimental settings reflect a lack of robustness in the models.

By inspecting gradients for each variable, we observe that age is predominant, followed by GOT, O2 saturation, PCR, Glucose, Platelets of secondary importance, in agreement with expert knowledge. This is reflected in our method, where our model attains peak performance when fusing the image modality with 7 clinical variables. Furthermore, our DMIB maintains peak performance upon adding more noisy and redundant clinical features, unlike most existing methods which show declining trend.  Another noteworthy point is when fusing the CT modality with a random noise input, our method retains its performance, whereas all existing fusion methods except concatenation method experience considerable performance drops. Again, this demonstrates the robustness of our approach and its ability to filter out noises. Overall, our method is more feasible and reliable for real-world clinical applications, when noise and redundant information are often present, and there is no prior guidance for which features are informative.

We discuss some insights for the various fusion schemes. Concatenation \cite{meng2020deep} fails to surpass the clinical modality for 1 and 4 clinical variables. This is probably due to the image feature dimensions being much larger than the clinical feature dimensions, resulting in the dominant influence of one modality. Channel-wise attention fusion \cite{duanmu2020prediction} only boosts performance when the modality data is complete and informative. It is particularly sensitive to noise, and drops significantly upon inclusion of noisy and redundant data. Transformer-based fusion also delivers very poor performance, never surpassing single-modality results. This is consistent with observations that attention-based mechanisms and transformers typically require a larger corpus of training data for effective training \cite{xu2022multimodal}. Dynamic fusion \cite{han2022multimodal}, subspace-based fusion \cite{zhou2021cohesive} and graph-based fusion \cite{mohla2020fusatnet} demonstrated better consistency and solid fusion performance. However, they are prone to be affected by noisy modalities. For subspace projection-based fusion, replacing the clinical variable with noise leads to considerable performance dips. Intuitively, aligning normal inputs to random noise could remove predictive information in the normal channels. For dynamic fusion, learning to weigh the reliability of features from a modality does not generalize well to test samples at an instance level, leading to a drop when including unreliable clinical inputs.

\noindent \textbf{iCTCF.} 
Furthermore, we demonstrate the exceptional fusion performance of DMIB when combined with all clinical data in both prognosis and diagnosis tasks, along with its resilience to noise input, as evidenced in Table~\ref{tab:iCTCF} and Table~\ref{tab:iCTCFdiagnosis}. 
Notably, when compared to using CT only, adding the singular feature `Age', has a significant impact on predicting COVID-19 patient outcomes (Table~\ref{tab:iCTCF}). However, it shows only a marginal improvement in the COVID-19 diagnosis task (Table~\ref{tab:iCTCFdiagnosis}).
As such, for COVID19 diagnosis, age can be deemed as a redundant feature and indeed. We observe similar performance in iCTCF when combining the CT modality with the noise feature. In both tables, our method and the concatenation method demonstrate the highest level of  robustness, with our method achieving the best fusion
performance.

\noindent \textbf{BRCA \& ROSMAP.}
As reported in Table \ref{tab:BRCAROSMAP}, DMIB is slightly inferior for breast cancer subtype classification but significantly superior for Alzheimer's diagnosis than \cite{han2022multimodal}, which  proves its adaptability  across different modalities.

\subsection{Ablation studies}
We performed ablation studies on the ITAC and BRCA datasets to study the effectiveness of various key components, with results reported in Table \ref{tab:ablation}. It is observed that preserving all information (No. 1) is worse than incorporating a bottleneck with sufficiency loss (No. 4, 5, 6). Furthermore, employing a bottleneck module without the sufficiency loss also fails to be useful, which is consistent with our intuitions that unconstrained discarding of information could remove predictive information (comparing No. 2, 3 to No. 6). Moreover, including modal-specific supervision $\L_{\mbox{\small{modality}}}$ leads to improvement, suggesting that boosting the extraction of modality-level task-relevant information facilitates learning of fused feature (No. 4 and No. 6). This provides strong evidence for the effectiveness of the sufficiency loss in preserving predictive information.

We further investigate whether DMIB remains effective under different network architectures by switching the image backbones for the prognosis experiment on our ITAC dataset. As reported in Table~\ref{tab:backbones}, we experimented with ResNet18, ResNet50 \cite{he2016deep}, EfficientNet \cite{tan2019efficientnet} and PoolFormer \cite{yu2022metaformer}. Regardless of the backbone, DMIB performs significantly better than the baselines of CT only and fusion methods such as concatenation (which retains all information) and subspace-based fusion (which extracts common information). These results strongly affirm the efficacy of our approach and its viability as a plug-and-play component for multimodal classification tasks.

\section{Conclusion}

Utilising insights from mutual information and information bottleneck theory, we have introduced a general multimodal classification approach which has achieved a state-of-the-art performance in various clinical tasks. Notably, our method holds great significance and applicability in the field of clinical practice, as it consistently delivers high performance, even when faced with limited training data and noisy input modalities. In the context of future research, our efforts will be directed towards expanding the application of information bottleneck theory to facilitate feature importance analysis and enhance the interpretability of multimodal learning.

\section*{Acknowledgment}
This study was supported in part by the ERC IMI (101005122), the H2020 (952172), the MRC (MC/PC/21013), the Royal Society (IEC\textbackslash NSFC\textbackslash 211235), the NVIDIA Academic Hardware Grant Program, the SABER project supported by Boehringer Ingelheim Ltd, Wellcome Leap Dynamic Resilience, and the UKRI Future Leaders Fellowship (MR/V023799/1).

\clearpage
\appendix
\section{Preliminary Definitions}
\noindent Given continuous random variables $X,Y,Z$, supported on $\mathcal{X},\mathcal{Y},\mathcal{Z}$ with probability distributions $p_X,p_Y,p_Z$: \\


\noindent $\left(\romannumeral1\right)$ The definition of mutual information of $X$ and $Y$ and its relation to information entropy:
\begin{equation}
\small
\begin{aligned}\label{eqn:MI}
I(X ; Y) & \equiv \mathbb{E} \left[ \log \frac{p_{X, Y}(X, Y)}{p_X(X) p_Y(Y)}\right]\\
&=\int_{\mathcal{X},\mathcal{Y}} p_{X, Y}(x, y) \log \frac{p_{X, Y}(x, y)}{p_X(x) p_Y(y)}  dx dy \\
&= \mathbb{E} [\log p_{X, Y}(X, Y)] - \mathbb{E}[p_X(X)] -  \mathbb{E}[p_Y(Y)] \\ 
&= - H(X,Y) + H(X)+  H(Y) \\
&= H(Y)-H(Y|X) \\
&= H(X)-H(X|Y)
\end{aligned}
\end{equation}~\\
where
\begin{equation*}
\begin{aligned}
H(X) & \equiv \mathbb{E} [-\log p_X(X)] = -\int_\mathcal{X} p_X(x) \log p_X(x) dx\\
H(X,Y) &\equiv \mathbb{E} [-\log p_{X,Y}(X,Y)]\\
&= -\int_{\mathcal{X},\mathcal{Y}}p_{X,Y}(x,y) \log p_{X,Y}(x,y) dx dy \\
H(Y|X) &\equiv \mathbb{E} [-\log p_{Y|X}(Y|X)] \\
&= -\int_{\mathcal{X},\mathcal{Y}}p_{X,Y}(x,y) \log p_{Y|X}(y|x) dx dy \\
\end{aligned}
\end{equation*}

\noindent (ii) The conditional mutual information of $X$ and $Y$ given $Z$ is defined as:
\begin{equation}
\small
\begin{aligned}
I(X;Y|Z)&\equiv \mathbb{E}\left[\log \frac{p_{X, Y|Z}(x, y|z)}{p_{X|Z}(x|z) p_{Y|Z}(y|z)}\right]\\
&=\int_{\mathcal{X},\mathcal{Y},\mathcal{Z}} p_{X, Y|Z}(x, y|z)p_Z(z) \\
&\quad\log \frac{p_{X, Y|Z}(x, y|z)}{p_{X|Z}(x|z) p_{Y|Z}(y|z)} dx dy dz
\end{aligned}
\end{equation}


\section{Proof of $I(f;f^*)=I(f;f^*|y) + I(y;f^*)$}
\begin{proof}
We have $I(f^*;y)= H(f^*) - H(f^*|y)$ by Eq.\eqref{eqn:MI}. Furthermore, since $f^*$ is obtained (deterministically) from $f$, we have $p_{f,f^*}(f,f^*)=p_f(f)$. Therefore, $I(f;f^*) =\mathbb{E}\left[-\log p_{f^*}(f^*)\right]=\mathbb{E}\left[\log\frac{p_{f,f^*}(f,f^*)}{p_{f^*}(f^*)p_{f}(f)}\right]= H(f^*)$ and similarly, $I(f;f^*|y) = H(f^*|y)$. Combining these, we have the desired result.
\end{proof}

\begin{table*}[t]
\centering
\caption{Clinical variables in ITAC and percentage of missing data}\label{tab:clinical}
\resizebox{1\textwidth}{!}{
\begin{tabular}{|l|l|l|l|l|l|l|l|l|l|}
\hline
$\#$ & 1 & 2 & 3 & 4 & 5 & 6 & 7 & 8 & 9 \\\hline
Variable & Age & Oxygen saturation & Platelets & Measured saturation oxygen & Respiratory rate &  PO2 &  D-Dimer & Cough & Dyspnea \\ \hline
Missing (\%) & 0 & 7.60 & 4.42 & 35.69 & 42.76 & 31.45 & 44.52 & N/A & N/A \\ \hline
$\#$ & 10 & 11 & 12 & 13 & 14 & 15 & 16 & 17 & 18 \\\hline
Variable & Diabetes & Neurological disease & Other CV disease & Admitted to ICU & Glucose &Urea & eGFR & GOT & PCR  \\ \hline
Missing (\%) & N/A & N/A & N/A & N/A & 15.72 & 7.24 & 7.77 & 24.56 & 16.43\\ \hline
\end{tabular}}
\end{table*}

\section{Proof of Proposition}
$KL\left[p(y|f) \| p(y|f^*)\right]=0 \implies I(y;f)-I\left(y;f^*\right)=0$
\begin{proof}
\begin{equation*}
\begin{aligned}
&  I(y;f)-I(y;f^*) = \\
&-\int p\left(f^*\right) p\left(y|f^*\right) \log p(y|f^*) df^* dy \\
&\;\;\;\; + \int p(f) p(y|f) \log p(y|f) df dy\\
&=-\int p\left(f^*\right) p\left(y | f^*\right) \log \left[\frac{p\left(y | f^*\right)}{p(y | f)} p(y | f)\right] d f^* d y\\
&\;\;\;\;+\int p(f) p(y | f) \log \left[\frac{p(y | f)}{p\left(y | f^*\right)}p\left(y | f^*\right)\right] d f d y\\
&= -\int p\left(f^*\right) KL\left[p(y | f^*) \| p\left(y | f\right)\right] d f^* \\
&\;\;\;\;-\int  p\left(f^*\right) p\left(y | f^*\right) \log p(y | f) df^*d y \\
&\;\;\;\;+\int p(f) KL[p(y | f) \| p(y | f^*)] d f\\
&\;\;\;\;+\int  p\left(f\right) p\left(y | f\right) \log p\left(y | f^*\right) df dy\\
&=\mathbb{E}_{f}\left[KL[p(y | f) \| p(y | f^*)]\right] - \mathbb{E}_{f^*}\left[KL[p(y | f^*) \| p(y | f)]\right] \\
&\;\;\;\;+ \int p(y) \log \frac{p(y | f^*)}{p\left(y|f\right)} dy\\
&\leq \mathbb{E}_{f}\left[KL[p(y | f) \| p(y | f^*)]\right] + \int p(y) \log \frac{p(y | f^*)}{p\left(y|f\right)} dy.\end{aligned}
\end{equation*}

Using Jensen's inequality and the fact that $-\log$ is strictly convex, we can show that the KL-divergence is always non-negative and the equality only holds when the distributions are equal almost-everywhere, which is proven as below:
\begin{equation}
\begin{aligned}\label{jensen}
KL[P\|Q] &= \mathbb{E}\left[-\log \frac{Q}{P}\right]& \\
&\geq -\log \mathbb{E} \left[\frac{Q}{P}\right] \;(\text{by Jensen's inequality})& \\
&=-\log\int_\mathcal{X} \frac{Q(x)}{P(x)}P(x)dx=0&
\end{aligned}
\end{equation}
where $P$ and $Q$ are two arbitrary distributions supported on $\mathcal{X}$. We have $KL[P\|Q] \geq 0$.

Hence, when $KL\left[p(y|f) \| p(y|f^*)\right]=0$, we have $p\left(y | f^*\right) = p\left(y | f \right)$ almost everywhere (follows from Eq. \eqref{jensen}), which implies $\int p(y) \log \frac{p(y | f^*)}{p\left(y|f\right)} dy=0$ and hence $I(y;f)-I(y;f^*)\leq 0$.  We also have $ I(y;f)-I(y;f^*) \le 0$, therefore $KL\left[p(y|f) \| p(y|f^*)\right]=0 \implies I(y;f)-I\left(y;f^*\right)=0$.
\end{proof}

\section{Summary of clinical variables in ITAC}
The overview of the missing data in the clinical variables in ITAC is given in Table \ref{tab:clinical}. We simply fill the missing value by the mean value calculated from the overall datasets.

\clearpage
\bibliographystyle{ieee_fullname}
\bibliography{References}
\end{document}